\documentclass[11pt]{article}

\usepackage{amssymb,amsmath,latexsym, hyperref, cite, natbib, graphicx, refstyle, pgfplotstable, booktabs, rotating, color, tabularx, authblk, array, url}

\usepackage{multirow}
\newcolumntype{C}[1]{>{\centering\arraybackslash}p{#1}}

\usepackage{geometry}
\usepackage{setspace}
\usepackage[multiple, hang,splitrule]{footmisc}


\usepackage[font=small, margin=0cm]{caption}

\begin{document}
\title{\textbf{Strategic Intelligence on Emerging Technologies: Scientometric Overlay Mapping}}

\author[1]{\textbf{Daniele Rotolo}\thanks{Corresponding author: d.rotolo@sussex.ac.uk, @danielerotolo, Phone: +44 1273 872980}}
\author[2,1]{\textbf{Ismael Rafols}\thanks{i.rafols@ingenio.upv.es}}
\author[1]{\textbf{Michael Hopkins}\thanks{m.m.hopkins@sussex.ac.uk}}
\author[3]{\textbf{Loet Leydesdorff}\thanks{loet@leydesdorff.net}}

\affil[1]{\small SPRU --- Science Policy Research Unit, University of Sussex, Brighton, United Kingdom}
\affil[2]{\small Ingenio (CSIC-UPV), Universitat Politecnica de Valencia, Valencia, Spain}
\affil[3]{\small Amsterdam School of Communication Research (ASCoR), University of Amsterdam, Amsterdam, The Netherlands}

\date{Version: \today \linebreak 
Accepted for publication in the \linebreak\textit{\textbf{Journal of the Association for Information Science and Technology}}\thanks{\href{http://dx.doi.org/10.1002/asi.23631}{{\color{blue}DOI: 10.1002/asi.23631}}. \copyright 2015 Rotolo, Rafols, Hopkins, Leydesdorff. Distributed under \href{http://creativecommons.org/licenses/by/4.0/}{{\color{blue}CC-BY-NC-ND}}.}}

\maketitle
\begin{abstract}
\onehalfspacing
\onehalfspacing
\noindent This paper examines the use of scientometric overlay mapping as a tool of 'strategic intelligence' to aid the governance of emerging technologies. We develop an integrative synthesis of different overlay mapping techniques and associated perspectives on technological emergence across the geographical, social, and cognitive spaces. To do so, we  longitudinally analyse (with publication and patent data) three case-studies of emerging technologies in the medical domain. These are: RNA interference (RNAi), Human Papilloma Virus (HPV) testing technologies for cervical cancer, and Thiopurine Methyltransferase (TPMT) genetic testing. Given the flexibility (i.e.\ adaptability to different sources of data) and granularity (i.e.\ applicability across multiple levels of data aggregation) of overlay mapping techniques, we argue that these techniques can favour the integration and comparison of results from different contexts and cases, thus potentially functioning as platform for a 'distributed' strategic intelligence for analysts and decision-makers.\newline\newline
\par
\end{abstract}
\clearpage 

\section{Introduction}
Emerging technologies are radically novel and relatively fast growing technologies characterised by a certain degree of coherence and with the potential to exert a considerable socio-economic impact \citep[]{Rotolo2015}. This impact is conceived in term of changing of existing industries \citep{Day2000}, the basis of competition \citep{Hung2006}, and human understanding or capabilities \citep{Alexander2012} and it is understood to be long term (typically with a 15-year horizon or so) \citep{Porter2002}. Change to the \textit{status quo} promised by emerging technologies, however, is somewhat uncertain and ambiguous since it lies in the future \citep{Rotolo2015}. The directionality of emerging technologies is the result of a variety of factors including visions and expectations of the actors involved \citep[e.g.][]{vanLente1998,Stirling2007, Collingridge1980}. Emergence is of a reflexive nature: actors are at the same time regulated by and regulating the emergence process. 

For these reasons, the governance of emerging technologies has assumed an increasing relevance, becoming a priority and part of the research agenda of many national governments. Developing policies to stimulate emergence towards societal benefit as well as to reduce the risk of the unintended effects or undesirable uses of emerging technologies is, however, a difficult activity. The governance of emerging technologies also assumes an emergent character resulting from the multitude of interactions of involved actors whereby the explicit attempts to shape arrangements are only one part of this process \citep[e.g.][]{Kooiman1993,Verbong2007}. Governance is the result of both intentional and un-intentional influences, \textit{de facto} including factors and actors non directly or purposefully aiming at governing irreversibilities \citep{Rip2010,David1985}. The risk that technological alternatives may be crowded out adds to this complexity \citep{Martin1995}.

In this context, the governing of emerging technologies requires tools of 'strategic intelligence', i.e.\ tools that are able to provide 'intelligence' inputs to the decision-making process for the development of policy instruments capable to cope with the rapid, uncertain, and ambiguous evolvements of technological emergence \citep{Kuhlmann1999, Stirling2007}. Conventional tools of strategic intelligence include science and technology foresight, innovation policy evaluation, and technology assessment. Frameworks on the \textit{use for strategic intelligence} of scientometrics in the context of emerging technologies are, however, less explored despite the large variety of scientometric studies developing approaches for the detection and the analysis of emerging technologies \citep[e.g.][]{Small2014,Glanzel2012,Porter1995}. 

The present paper aims to fill this gap by focusing on a recently developed scientometric mapping approach, namely \textit{overlay mapping} \citep[see][]{Rafols2010}. This approach has increasingly attracted the attention of scientometric scholars since its capability to convey relatively complex information about technological emergence through maps the interpretation of which does not require advanced scientometric skills \citep[e.g.][]{Klavans2009,Leydesdorff2013j,Waltman2012,Leydesdorff2014}. This is achieved by projecting an \textit{overlay} of data representing a focal emerging technology over a \textit{basemap}, which represents the geographical, social, and cognitive (or combinations of those) spaces of emergence. For example, the dynamics of an emerging technology can be traced by projecting longitudinally publications and patents associated with the technology over basemap depicting the wider activity in science and technology (at different levels such as journals, patent classes, research domains) or geographical areas.

To examine the extent to which overlay mapping can aid the governance of emerging technologies by functioning as a tool of strategic intelligence, we analyse three case-studies of emerging technologies in the medical domain: (i) RNA interference (RNAi), (ii) cervical cancer testing technologies for Human Papilloma Virus (HPV) and (iii) genetic testing for the Thiopurine Methyltransferase (TPMT) enzyme. These are used as 'vignettes' to generate an integrative synthesis of the 'intelligence' that different overlay mapping approaches can generate on technological emergence. For example, once suitable source data are available, analysts and decision-makers can gather (with relatively low effort thanks to the assistance of publicly available software routines) 'intelligence inputs' such as: urban areas producing the large amount of knowledge as well as the knowledge that is more likely to be of higher future impact; geographical distributions of key organisational actors in collaborative networks; and identification of the main scientific disciplines and technological domains involved in the emergence including dynamics. In addition, overlay mapping techniques provide analysts and decision-makers with flexibility and granularity in the analytical process. 

These tools are flexible in the sense that they are not constrained to the institutional boundaries of the databases where relevant data are maintained \citep{Griliches1994}. Perspectives on the emergence can be built by using same or similar search strings across multiple sources of data. Different levels of granularity can be obtained providing the possibility of multiple levels of analysis of the emergence process in order to reveal macro-, meso-, and micro-dynamics. Flexibility and diverse levels of granularity also favour the integration and comparison of results from different contexts. Overlay mapping may also suggest to analysts and decision-makers new directions of investigation as well as feeding into political discourse in a timely manner.

The paper is structured as follows. In the next section, we introduce the overlay mapping approach and the diverse possibilities of applications it provides for the analysis of emerging technologies. We then provide some background to the selected three case-studies and introduce used data sources in Section 3. This section also discusses an important caveat associated with the analysis of emerging technologies, that is the delineation of the boundary of \textit{what is emerging}. Results of overlay mapping applied to the three case-studies are presented in Section 4. The use of overlay mapping for strategic intelligence is then discussed in Section 5. Section 6 concludes by highlighting the contribution of the paper.

\section{Scientometric overlay mapping}
The scientometric community has made great efforts in the development of techniques to trace science and technology dynamics since the seminal works by \cite{Price1965}, \cite{Small1973}, \cite{Garfield1979}, and \cite{Callon1983}. Well established examples of these methodological approaches are direct citation and co-citation analyses \citep[e.g.][]{Small1977,Garfield1964}, bibliographic coupling \citep[e.g.][]{Kessler1963}, and co-words analysis \citep[e.g.][]{Callon1983}. A multitude of indicators, which mostly rely on publication and patent data, are based on this type of scientometric tools.

\begin{figure}[h]
\begin{center}
\includegraphics[width=9cm]{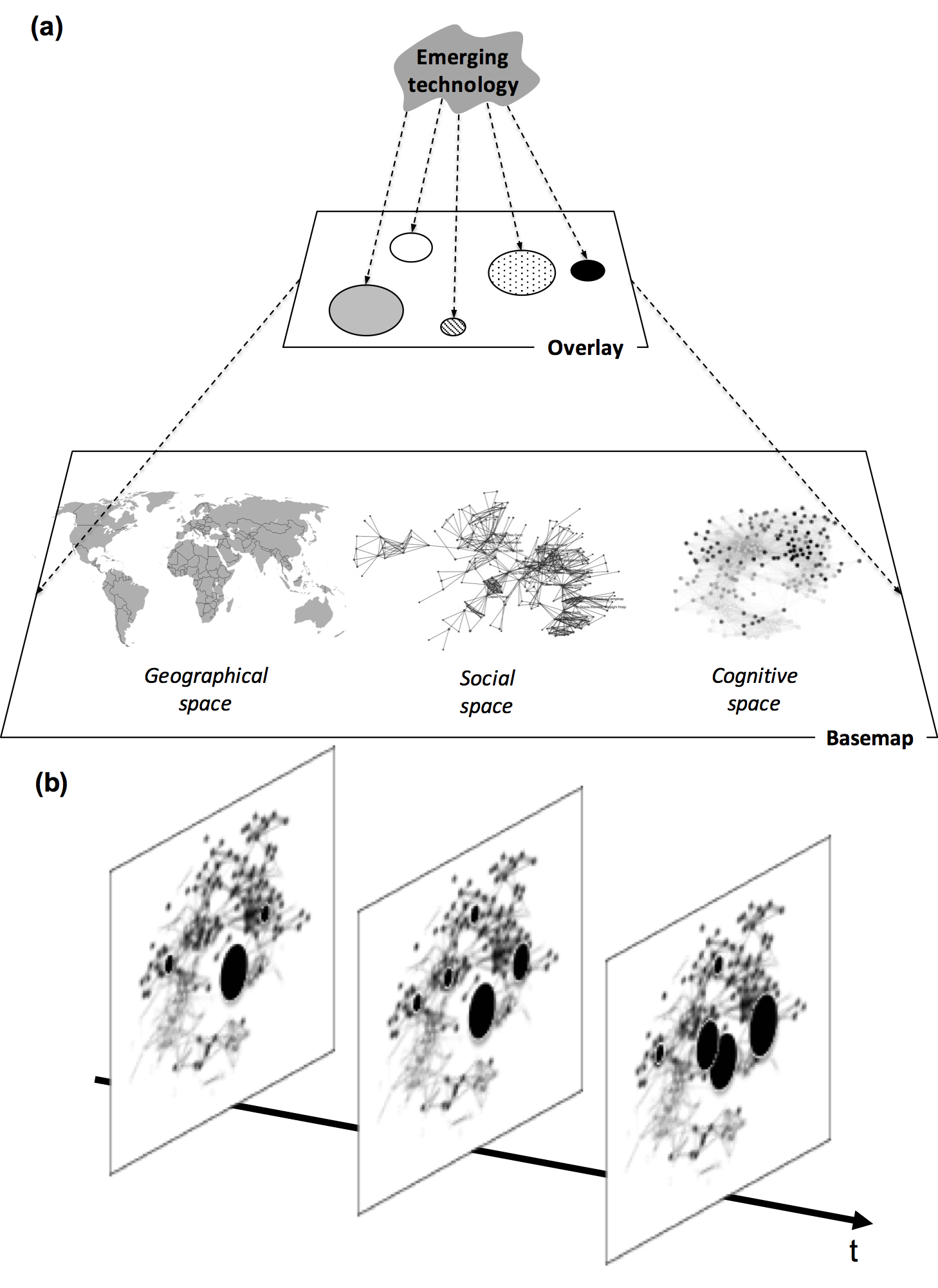}
\end{center}
\caption{Overlay mapping: basic principle, spaces of emergence (a) and longitudinal overlays (b). \newline\textit{Source: Authors' elaboration.}}
\label{fig:overlay}
\end{figure}

The growing attention to emerging technologies and their potential to change the \textit{status quo} has also led scientometric scholars to focus their efforts on methodological approaches for the detection and analysis of emergence in science and technology \citep[e.g.][]{Small2014,Glanzel2012,Porter1995}.\footnote{For a review of the scientometric contribution to the detection and analysis of emerging technologies see \cite{Rotolo2015}.} These efforts and seminal contributions on science and technology (S\&T) mapping for policy use \citep[e.g.][]{Noyons2001,Borner2010}, together with the increase in computational power, the improved performance of processing algorithms, and the higher technical capabilities in exploiting databases in a more comprehensive manner, have spurred the development of a number of mapping techniques among which \textit{overlay mapping} \citep[e.g.][]{Rafols2010}. The basic idea underlying this approach is depicted in \Figref{overlay}. \textit{Overlays} are used to project publication and patent data about an emerging technology over \textit{basemaps}, which depict the wider activity in science and technology, social structures, or geographical areas (\Figref{overlay}a).\footnote{It is worth noting that the application of overlay mapping is not limited to emerging technologies. It can be also applied to visualise the publishing and patenting activity of individual researchers, organizations (e.g.\ firms, universities, research institutes), or communities.} This process is iterated at different time periods to visualise dynamics (\Figref{overlay}b). 

A basemap can be conceived as a set of $n$ nodes, a $n \times n$ adjacency matrix $\mathbf{B}$ describing the linkages between these nodes, two vectors $\mathbf{x}$ and $\mathbf{y}$ containing 2D spatial coordinates of nodes, and a vector $\mathbf{c}$ including the information about which clusters nodes belong to: 

\[
\mathbf{B} =\left(\begin{array}{@{}cccc@{}}
     b_{11}&b_{12}&\cdots &b_{1n} \\
  b_{21}&b_{22}&\cdots &b_{2n} \\
  \vdots & \vdots & \ddots & \vdots\\
  b_{n1}&b_{n2}&\cdots &b_{nn}
\end{array}\right)
;
\mathbf{x}=\left(\begin{array}{@{}c@{}}
    x_1 \\
    x_2 \\
    \cdots \\
    x_n
\end{array}\right)
;
\mathbf{y}=\left(\begin{array}{@{}c@{}}
    y_1 \\
    y_2 \\
    \cdots \\
    y_n
\end{array}\right)
;
\mathbf{c}=\left(\begin{array}{@{}c@{}}
    c_1 \\
    c_2 \\
    \cdots \\
    c_n
\end{array}\right)
\]

\noindent Nodes can represent cities, authors/inventors, organisations, or knowledge areas (at different levels such as journals, patent classes, disciplines, topics), while the value of the linkage $b_{ij}$ between nodes $i$ and $j$ (for $i,j=1,2,\dots, n$ and $i \neq j$) can refer to social structures (e.g.\ co-authorships), indicate similarity between nodes, or assume value zero in the case of simple geographical basemaps. It is worth noting that the adjacency matrix is symmetric ($b_{ij}=b_{ji}$) and the diagonal is not used for overlay mapping visualisation purposes, thus it can be assumed $b_{ii}=0$. The 2D spatial coordinates of nodes are defined on the basis of the layout algorithm used to visualise the basemap. The information of how nodes are clustered is often used in the case of cognitive basemaps to identify different knowledge domains (e.g.\ disciplines). 

Previous research has generated a number of basemaps. These include basemaps of Web of Science (WoS) Categories \citep{Rafols2010}, journals \citep{Leydesdorff2013j}, medical terms (namely, Medical Subject Headings (MeSH) descriptors) \citep{Leydesdorff2012}, and patent technological classes \citep[e.g.][]{Kay2014,Leydesdorff2014}. Google Maps have been instead used as geographical basemaps \citep{Leydesdorff2010,Bornmann2011,Leydesdorff2012a}, while basemaps representing 'global' social structures have been less explored.

The set of longitudinal overlays can be described as an array of vectors of $n$ elements:

\[
\mathbf{o_1}=\left(\begin{array}{@{}c@{}}
o_{11}\\
o_{21}\\
\vdots\\
o_{n1}
\end{array}\right),
\mathbf{o_2}=\left(\begin{array}{@{}c@{}}
o_{12}\\
o_{22}\\
\vdots\\
o_{n2}
\end{array}\right),
\cdots, 
\mathbf{o_t}=\left(\begin{array}{@{}c@{}}
o_{1t}\\
o_{2t}\\
\vdots\\
o_{nt}
\end{array}\right),
\cdots,
\mathbf{o_T}=\left(\begin{array}{@{}c@{}}
o_{1T}\\
o_{2T}\\
\vdots\\
o_{nT}
\end{array}\right)
\]
\noindent where $o_{it}$ indicates the value that the node $i$ of the selected basemap assumes at time period $t$ (for $i,=1,2,\dots, n$ and $t=1,2,\dots, T$). This, for example, can indicate the number of publications at time period $t$ that fall within the WoS Category $i$, are published in the journal $i$, were produced by authors based in the city $i$, etc. These overlays are then projected over the basemap in terms of size of nodes, while the spatial coordinates of nodes $\mathbf{x}$ and $\mathbf{y}$ and (if any) clustering $\mathbf{c}$ are kept constant. This process can be carried out by analysts and decision-makers with relatively low efforts due to the existence of publicly available software routines. Also, the resulting maps do not require advanced scientometric skills for their interpretations --- notwithstanding the importance of interacting with experts of the considered case of emerging technology. 

The evolutionary perspectives generated with overlay mapping can potentially reveal \textit{de facto} (intentional and unintentional) government arrangements among factors and actors involved in the emergence process. These perspectives may facilitate timely analyses on relevant dynamics of emergence (e.g.\ main geographical areas involved in the knowledge production processes and collaborative interactions among those, main actors in inter-organisational networks and their geographical distribution, domains of science and technologies involved in the emergence and associated dynamics across disciplines, journals, technological classes, and, in the case of medical technologies, medical areas), thus complementing the existing set of more conventional tools of strategic intelligence for the policy-making of emerging technologies. This, in turn, may support the development of policy instruments for governance that are more capable to cope with the rapid dynamics of emergence and associated uncertainty and ambiguity. 

Flexibility in the use of diverse sources of data and multiple levels of granularity in the analysis are additional features that make overlay mappings a useful tool for a 'distributed' strategic intelligence on emerging technologies. Results from different contexts can reflexively and discursively be integrated and compared with relatively little additional effort --- basemaps provide common ground for comparisons. The choice of overlays and basemaps, however, depends on the specific questions the analyst or decision-maker aims to address. 

We build on the studies mentioned above to deepen the examination of the use of overlay mapping as a tool of strategic intelligence for the governance of emerging technologies. It is worth noting that reiterating the full technical details of developed overlay mapping approaches is not within the scope of the present paper --- the reader is referred to the corresponding studies, which provide webpages and software routines. We instead aim to provide an integrative synthesis of different overlay mapping techniques and the types of perspectives on emergence these enable analysts and decision-makers to generate. To do so, we will apply overlay mapping techniques to three case-studies of emerging technologies.

\section{Methods}
\subsection{Background on the case-studies}
The case-studies of emerging technologies in the medical domain we examine with overlay mapping are: (i) RNAi, (ii) HPV testing technologies for cervical cancer, and (iii) diagnostic technologies for the TPMT enzyme. These case-studies exhibit diversity in terms of their scale, breadth of applications and position with regard to alternative technological options. Thus, from a governance perspective, each technology raises different challenges: RNAi is a technology with radical potential across the domains of therapeutics, diagnostics and as a basic research tool, but with fluctuating industrial interest due to technical challenges in delivering on its promised applications; HPV testing is battling the entrenched technology for cervical cancer screening, i.e. the Pap testing, with powerful interest groups on both sides; TPMT testing technologies are a much smaller technological field, exploited in a series of small clinical niches, mainly in developed countries, where there are concerns that best practice may not be spreading. This diversity of case-studies provides us with the opportunity to discuss the use of overlay mapping to generate 'intelligence' inputs for the governance of emerging technologies in different contexts. However, at the same time, the three case-studies enable us to achieve a certain degree of external validity of the findings of our analyses \citep{Yin2003,Gibbert2008}. The comparison is somewhat controlled since these case-studies are from the medical domain and share primarily North American origin and similar periods of emergence, in the 1980s and 1990s. One would expect very different dynamics in other contexts such as electronic engineering --- and possibly more difficulties in tracing the technologies with publications and patents here or in contexts where English language publications are less of a focus for knowledge producers. In the following, we provide background sketches for the case-studies to contextualise the scientometric overlay maps presented later in the paper.

Firstly, RNAi is a molecular process that can silence the expression of genes. Gene regulation plays a critical role in the progression of cancers, genetic diseases, and infectious agents. By silencing specific genes one can stop the progression of some diseases. This 'small RNA' silencing mechanism was discovered in 1998.\footnote{Andrew Z. Fire and Craig C. Mello were awarded the 2006 Nobel Prize in 'Physiology or Medicine' for this discovery.} Its discovery reshaped the landscape of research on gene expression creating important expectations especially for the therapeutic applications \citep{Sung2006,Leydesdorff2011}.

Secondly, HPV testing technologies are positioned within a specific domain of application, i.e.\ the detection of HPV virus, which is recognised as causing cervical cancer.\footnote{Harald zur Hausen \citeyear{ZurHausen1987}, who discovered the association between the HPV and cervical cancer, was later awarded the 2008 Nobel Prize in 'Physiology or Medicine'.} Cervical cancer has a significant disease burden. About 528,000 new cervical cancers occur and cause about 266,000 deaths worldwide each year --- 85\% and 88\% of new cases and deaths affect women in less developed regions \footnote{GLOBOCAN 2012 available at \url{http://globocan.iarc.fr}.} This has led to the development of a large screening program. In the US more than 100 million tests are performed annually.\footnote{"Health, United States, 2012 - With Special Feature on Emerging Care", U.S. Department of Health and Human Services. Available at \url{www.cdc.gov/nchs/data/hus/hus12.pdf}.} While cervical cancer screening has been conducted for more than 50 years using the (cytology-based) Pap test to detect cancerous cells or cells potentially evolving into cancerous states, the discovery of the strong association between HPV infections (especially HPV types 16 and 18) and cervical cancer in the 1980s opened the space for the development of a competing and more sensitive testing technology based on molecular biology \citep{Hogarth2012a,Hogarth2015}.

Thirdly, similarly to HPV testing, TPMT testing technologies are positioned close to the applied-research domain. Yet, the clinical utility of TPMT's application has been contested across medical fields with different clinical guidelines supporting and discouraging the use of the test. TPMT testing is one of an emerging class of 'pharmacogenetic tests' which predict adverse events associated with pharmaceutical use in individuals that cannot metabolize a drug due to a genetic mutation \citep{Hopkins2006}. TPMT is an enzyme in the human body responsible for metabolising (i.e.\ breaking down) thiopurine drugs. Cytotoxic thiopurine drugs such as Azathioprine are used to treat a range of conditions including leukaemia, and autoimmune diseases (such as Lupus or rheumatoid arthritis). However, when a patient has mutations in the gene encoding TPMT, she/he is unable to metabolise the drug and is at increased risk of toxicity from a build-up of thiopurines. Following the discovery of genetic variations affecting drug metabolism, several types of TPMT testing technologies began to emerge in the 1990s with applications across a number of clinical fields of use such as oncology, dermatology, gastroenterology, and rheumatology.

\subsection{Data sources}
Overlay mapping mainly relies on two sources of data: publications and patents. We queried Thomson-Reuter's Web of Science (WoS) and MEDLINE/PubMed of the US National Library of Medicine (NLM) for publication data, and the United States Patent and Trademark Office (USPTO) for patent data. Data for each case-study were gathered with a set of keywords we identified by triangulating interviews with experts and quantitative research works on the case-studies (see \Tabref{data}). The interviews with experts in the fields of the selected case-studies are also used as lens of interpretations of the results we obtain from the application of overlay mapping approaches.

\setlength{\tabcolsep}{10pt}
\renewcommand{\arraystretch}{1.2}
\begin{table}\footnotesize
	\caption{\label{tab:data}Data sources and search strings.}
	\centering
\begin{tabular}{lllp{6.5cm}}
\hline\hline
Case-study   & Data         & Database       & Search string   \\
\hline
RNAi         & Publications & ISI WoS        & TI=siRNA or TI=RNAi or TI="RNA interference" or TI="interference RNA"\\
             &              &                &                                                                      \\
             &              & MEDLINE/PubMed & siRNA[Title] or RNAi[Title] or "RNA interference"[Title] or "interference RNA" [Title]          \\
             &              &                &                                                                                                           \\
             & Patents      & USPTO          & ACLM/(siRNA or RNAi or "RNA interference" or "interference RNA")                                                                                                     \\
\hline
HPV testing  & Publications & ISI WoS        & (TI=HPV* or TI="Human Papilloma Virus*" or TI="Human Papillomavirus*" or TI="Human Papilloma*virus*") and  (TI=Cervical or TI=Cervix) and (TI=diagnos* or TI=test* or TI=assay or TI=detect* or TI=screen* or TI=predict*)                                                         \\
             &              &                &                                                                                                                         \\
             &              & MEDLINE/PubMed & (HPV*[Title] or "Human Papilloma Virus*"[Title] or "Human Papillomavirus*"[Title]) and (Cervical[Title] or Cervix[Title]) and (diagnos*[Title] or test*[Title] or assay[Title] or detect*[Title] or screen*[Title] or predict*[Title]) \\
             &              &                &                                                                                                                     \\
             & Patents      & USPTO          & ACLM/((HPV or "Human Papilloma Virus\$" or "Human Papillomavirus\$") and (Cervical or Cervix) and (diagnos\$ or test\$ or assay or detect\$ or screen\$ or predict\$))                                                                                                                   \\
\hline
TPMT testing & Publications & ISI WoS        & TI=TPMT or TI= "Thiopurine Methyltransferase"                                                                                                                                \\
             &              &                &                                                                                                      \\
             &              & MEDLINE/PubMed & TPMT[Title] or "Thiopurine Methyltransferase"[Title]                                                                                                                             \\
             &              &                &                                                                                       \\
             & Patents      & USPTO          & ACLM/(TPMT or "Thiopurine Methyltransferase")  \\                                                                                                                                                                                                                                   
\hline\hline
\multicolumn{4}{l}{\footnotesize \textit{Source: Authors' elaboration based on interviews with experts and previous research works on the case-studies.}}
\end{tabular}
\end{table}

The retrieval of data for emerging technologies is challenging. Research topics often overlap and the vocabulary used to describe them is in flux \citep{Robinson2007a}. Most up-to-date methods for retrieval rely on combining lexical search and citation analysis \citep[e.g.][]{Zitt2006,Huang2015}, yet these approaches would be time-consuming for the non-expert analyst. We therefore deemed more suitable to use keyword-based searches given the parsimony and timeliness of the approach. We limited the search of keywords and their combinations in scientific articles' titles since this approach tends to generate less false positives as compared to the searching in articles' abstracts. Abstracts often contain technical and methodological terms not representing the core of knowledge the given article claims \citep{Leydesdorff1989}. An extended search may generally retrieve many additional records, yet with the risk of including many ones not closely related to the given emerging technology one aims to trace --- it increases 'recall' at the expenses of losing 'precision'. 

The identification of patents that are relevant for the building of informative perspectives on emerging technologies requires a different approach. The incentives to patent are indeed different from those underlying the publication of scientific articles. The primary purpose of the patent system is to reward patentees by providing them a temporary monopoly to commercially exploit the patented inventions. This requires patentees to disclose the technical knowledge of the inventions. For this reason, valuable information tends to be included in patent claims that inform about the scope of the technical knowledge of the considered invention \citep{Hunt2007} --- claims "define the invention and are what aspects are legally enforceable" (USPTO Glossary).\footnote{The USPTO Glossary is available at \url{www.uspto.gov/main/glossary}.}  We therefore focused the search of keywords on patent claims. Issues related to the definition and delineation of the boundaries of emerging technologies (e.g.\ the identification of an effective set of keywords, limited data coverage) are further discussed in the next section.

The number of publications and patents that relate to each of the three case-studies, from 1982 to 2011, are reported in \Tabref{cases}.\footnote{Data on USPTO patent applications are accessible since year 2001. The filing year of patents was considered.} While the first publications related to TPMT and HPV stem from the early 1980s, data for RNAi appears only since 1998 when this silencing mechanism was discovered and first published. Publication data from WoS and MEDLINE/PubMed allow for relatively simple but informative reports on the emergence of these three technologies in terms of published scientific articles. Growth can be observed for the three case-studies. Yet, it is clear that the pace of this growth as well as the scale of this emergence is significantly different from one case to another in two respects. First, the growth in the number of publications for RNAi is steeper than the other two case-studies. Second, RNAi and HPV testing technology show an increasing number of publications for the entire observation period. Conversely, the testing technology for TPMT enzyme seems to have reached its mature phase, in terms of volume of publications, in the last few years of observation.

\setlength{\tabcolsep}{5pt}
\renewcommand{\arraystretch}{1.2}
\begin{sidewaystable}\scriptsize
	\caption{\label{tab:cases}Publication and patent data for the three case-studies.}
	\centering
\begin{tabular}{cccccccccccccccc}
\hline\hline
 & & \multicolumn{4}{c}{RNAi} & & \multicolumn{4}{c}{HPV testing}  & & \multicolumn{4}{c}{TPMT testing}  \\
  \cline{3-6}   \cline{8-11}   \cline{13-16}
Year & & ISI WoS & MEDLINE & USPTO & USPTO & & ISI WoS & MEDLINE & USPTO & USPTO & & ISI WoS & MEDLINE & USPTO & USPTO \\
 & & & PubMed & (granted) & (applications) & & & PubMed & (granted) & (applications)  & & & PubMed & (granted) & (applications)  \\
\hline
1982  &  &      &      &     &      &  & 0    & 0    & 0  & 0   &  & 3   & 2   & 0  & 0  \\
1983  &  &      &      &     &      &  & 2    & 1    & 0  & 0   &  & 2   & 0   & 0  & 0  \\
1984  &  &      &      &     &      &  & 0    & 0    & 0  & 0   &  & 0   & 2   & 0  & 0  \\
1985  &  &      &      &     &      &  & 4    & 2    & 0  & 0   &  & 3   & 2   & 0  & 0  \\
1986  &  &      &      &     &      &  & 12   & 10   & 0  & 0   &  & 5   & 1   & 0  & 0  \\
1987  &  &      &      &     &      &  & 22   & 14   & 4  & 0   &  & 5   & 4   & 0  & 0  \\
1988  &  &      &      &     &      &  & 17   & 16   & 0  & 0   &  & 2   & 0   & 0  & 0  \\
1989  &  &      &      &     &      &  & 30   & 24   & 3  & 0   &  & 1   & 1   & 0  & 0  \\
1990  &  &      &      &     &      &  & 32   & 28   & 0  & 0   &  & 2   & 2   & 0  & 0  \\
1991  &  &      &      &     &      &  & 36   & 31   & 1  & 0   &  & 6   & 5   & 0  & 0  \\
1992  &  &      &      &     &      &  & 41   & 49   & 2  & 0   &  & 5   & 5   & 0  & 0  \\
1993  &  &      &      &     &      &  & 18   & 27   & 1  & 0   &  & 11  & 9   & 0  & 0  \\
1994  &  &      &      &     &      &  & 29   & 30   & 4  & 0   &  & 8   & 5   & 1  & 0  \\
1995  &  &      &      &     &      &  & 24   & 28   & 5  & 0   &  & 17  & 14  & 1  & 0  \\
1996  &  &      &      &     &      &  & 35   & 32   & 1  & 0   &  & 11  & 8   & 0  & 0  \\
1997  &  &      &      &     &      &  & 33   & 28   & 2  & 0   &  & 15  & 10  & 0  & 0  \\
1998  &  & 3    & 2    & 1   & 0    &  & 37   & 33   & 1  & 0   &  & 20  & 15  & 1  & 0  \\
1999  &  & 15   & 10   & 1   & 1    &  & 41   & 35   & 4  & 0   &  & 21  & 11  & 0  & 0  \\
2000  &  & 42   & 33   & 4   & 1    &  & 43   & 41   & 4  & 0   &  & 24  & 15  & 0  & 0  \\
2001  &  & 56   & 52   & 5   & 22   &  & 50   & 52   & 1  & 5   &  & 28  & 19  & 3  & 6  \\
2002  &  & 166  & 126  & 25  & 113  &  & 54   & 43   & 2  & 13  &  & 43  & 25  & 1  & 2  \\
2003  &  & 427  & 280  & 45  & 400  &  & 69   & 59   & 9  & 29  &  & 42  & 30  & 1  & 3  \\
2004  &  & 785  & 523  & 83  & 685  &  & 59   & 59   & 8  & 19  &  & 36  & 23  & 1  & 2  \\
2005  &  & 892  & 681  & 86  & 583  &  & 105  & 95   & 6  & 21  &  & 40  & 19  & 1  & 4  \\
2006  &  & 932  & 782  & 124 & 684  &  & 94   & 86   & 3  & 24  &  & 39  & 28  & 2  & 6  \\
2007  &  & 1002 & 737  & 131 & 752  &  & 121  & 91   & 6  & 16  &  & 30  & 15  & 0  & 1  \\
2008  &  & 1032 & 827  & 106 & 772  &  & 134  & 106  & 5  & 21  &  & 26  & 15  & 4  & 9  \\
2009  &  & 1016 & 837  & 118 & 767  &  & 154  & 120  & 5  & 33  &  & 43  & 30  & 0  & 1  \\
2010  &  & 1029 & 904  & 89  & 761  &  & 121  & 99   & 0  & 18  &  & 24  & 19  & 0  & 2  \\
2011  &  & 1131 & 969  & 35  & 410  &  & 143  & 118  & 0  & 11  &  & 28  & 16  & 0  & 5  \\
\hline
Total &  & 8528 & 6763 & 853 & 5951 &  & 1560 & 1357 & 77 & 210 &  & 540 & 350 & 16 & 41\\
\hline\hline
\multicolumn{16}{l}{\footnotesize \textit{Source: Authors' elaboration on the basis of WoS, MEDLINE/PubMed and USPTO data collected in January 2013.}}
\end{tabular}
\end{sidewaystable}

Patent data reveal similar distinctive features. The production of patents (both granted patents and patent applications) related to RNAi, for example, is relatively greater than in the case of HPV and TPMT testing technologies. This is not surprising given the former's wider scope of potential applications, especially therapeutics where patenting is routine. The apparent decline of the patenting activity related to RNAi in the last two years of observation is likely to be due to the delay of 18 months between the patent filing data and the publication of the patent document on the USPTO. Although most patenting in RNAi occurs in small firms, the decision of some large pharmaceutical companies, including Merck, Roche and Pfizer, to shut down their R\&D units on RNAi may have also resulted in a decreased interest on RNAi and contributed to this decline \citep{Lundin2011}.

The patent application activity around HPV testing technology grows from 2002 to 2004 and then stabilises in the subsequent years with a peak of applications in 2009. The low number of granted patents and patent applications for TPMT testing technology does not allow clear trends in the production of technical knowledge to be deduced, other than to note this is a field that appears to have attracted less attention.

\subsection{Definition and delineations issues}
A key preliminary input for the use of mapping and overlay techniques is the identification of a \textit{corpus} --- in this case, publications and patents associated with the emerging technology one aims to map. This task is generally much more iterative than one initially expects. The lack of a shared terminology due to the different meanings and expected uses associated with the technology makes the identification of a suitable set of keywords particularly challenging. Institutional vocabularies (e.g.\ MeSH descriptors or patent classifications) may also not be particularly helpful. Relevant categories (e.g.\ new terms or classes) may have not been yet included. This becomes even more problematic for contemporary cases of emerging technologies \citep{Rotolo2015}, notwithstanding the challenges associated with terminology changes and differing levels of coverage in datasets \citep[see][]{ Hopkins2013}. 

The ambiguity of what a 'technology' is adds to the difficulty of the task. A technology can include both the knowledge underpinning method or processes to fulfil a purpose (e.g.\ understanding of gene silencing) as well as the network of practices and components (e.g.\ the molecules and techniques that result in gene silencing). The case of RNAi is apt to highlight these issues.

RNAi is the naturally occurring process in which gene expression is reduced as a result of the breakdown of messenger RNAs. This process can be triggered by naturally occurring (endogenous) molecules, called microRNAs (miRNAs), or by externally inserted (exogenous) molecules, called small interfering RNAs (siRNAs). Soon after the discovery of RNA interference by \cite{Fire1998}, siRNAs were recognised as valuable for gene silencing, both as tools for research and for therapeutic purposes. This led to a boom in public and private R\&D investments \citep{Haussecker2008, Haussecker2012}. In parallel, but with some significant delay, it was realised that miRNAs were not a marginal phenomenon, but played a major role in gene regulation, including abnormal down or up regulation in certain diseases caused or earmarked by anomalous gene expression, such as many cancers. 

In principle, then, to delineate the research boundary of RNAi one should include both siRNAs and miRNAs. However, it turns out that 'RNA interference' as a technology (not as phenomenon) has become mainly known as the human-induced, exogenous interference, which was developed for therapeutic purposes. We then have two potential definitions of RNAi, one covering the entire field and one excluding miRNA --- the latter, commonly adopted for the discussion of therapeutic applications.

\begin{figure}[h]
\includegraphics[width=14cm]{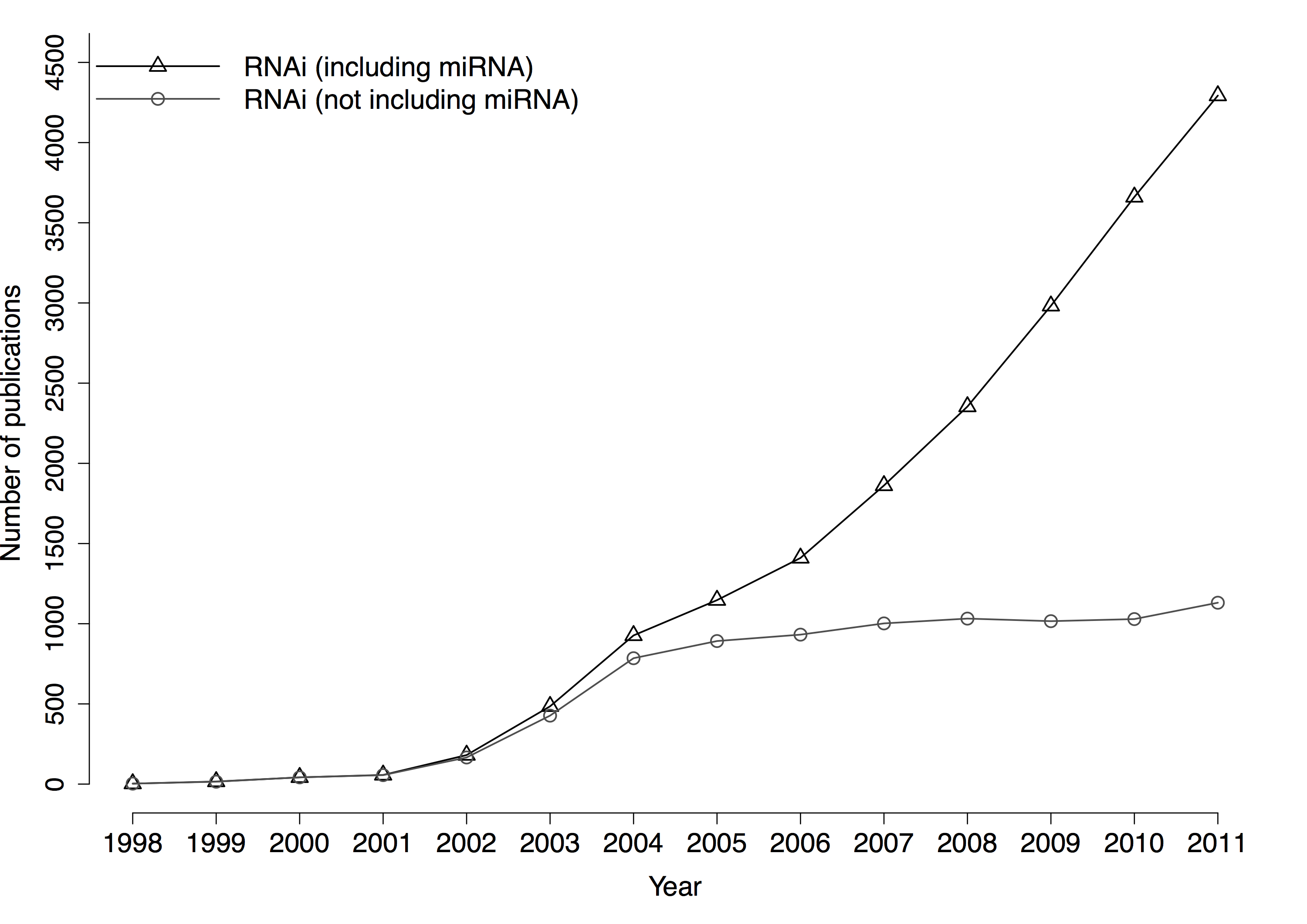}
\centering
\caption{Delineating the boundary of RNAi. Scientific articles related to miRNA were retrieved by using the following search string in WoS: 'TI=microRNA* or TI=miRNA*'. \newline\textit{Source: Authors' elaboration on the basis of WoS data.}}
\label{fig:delineation}
\end{figure}

\Figref{delineation} illustrates that the observed dynamics are different. Whereas miRNA research is still booming, possibly due to their potential use as biomarkers in diseases such as cancers, publications focused on siRNA have reached a plateau of about a thousand publications per year. The latter is possibly related to the challenges encountered in delivering siRNA in therapeutic applications that resulted in a retreat of pharmaceutical investment \citep{Haussecker2012}. These differences in trends support the views that the uses of miRNAs and siRNAs follow distinct trajectories, which it is best to differentiate. In this study, we focus on RNAi as research tool and for therapeutic applications, i.e.\ not including miRNA, which is not booming any more in terms of publications, but it is now beginning to reach use in some applications.

There is not an optimal solution to address the fuzziness of the boundaries of emerging technologies. The analyst often has to carry out a process that may take several iterations and require a significant interpretative input from experts in the considered technology field before producing a suitable set of keywords for the \textit{corpus} identification.

\section{Results}
The use of overlay mapping approaches as tools of strategic intelligence for policy-making processes is discussed below. We specifically build perspectives on the emergence process of the three selected case-studies across the geographical, social, and cognitive spaces. The mapping approach enables the analyst to create syntheses of relatively complex information on the dynamics of emerging technologies and to visualise these in maps. The maps generated can be easily interpreted and analysed with visual inspection (e.g.\ emergence of new cities involved in the production of publications and patents, intensification of the collaboration activity, involvement of new scientific disciplines or technological areas). For this reason, the results described below are mostly illustrative and complemented with the qualitative understanding of case-studies. 

However, it is worth noting that the aforementioned overlay mapping routines also allow the analyst to perform quantitative analyses. The overlay data are saved into a relational database, which can be queried to produce statistical information (for example, to perform hypothesis testing) on the focal emerging technology. As discussed, overlay data include: number of new countries/cities involved with the knowledge creation process, proportional increase of the number of publications in newly involved disciplines, and indexes of diversity of the scientific and technological areas involved. Due to space limitations, we discuss and report in this paper only a few illustrative maps from the results obtained --- we instead provide webpages to access the entire set of overlay maps.

\subsection{Tracing emergence in the geographical space}
The projection of overlays of publication and patent data over geographical (base)maps can visualise the emergence process across cities, regions, and nations. Building on previous work \citep[e.g.][]{Bornmann2011,Bornmann2011a, Hu2012, Leydesdorff2012a}, one can, for example, identify sites for a given emerging technology where highly-cited scientific articles were published more frequently than expected. This is depicted in terms of the sizes and colours of nodes --- Google Maps are used as  geographical (base)maps.\footnote{The $z-test$ for two independent proportions is used. The null hypothesis is the randomness in the selection of papers for a city \citep[see][]{Bornmann2011}. A threshold of top 10\% most-frequently cited scientific articles is selected.}

\begin{figure}[h]
\includegraphics[width=16cm]{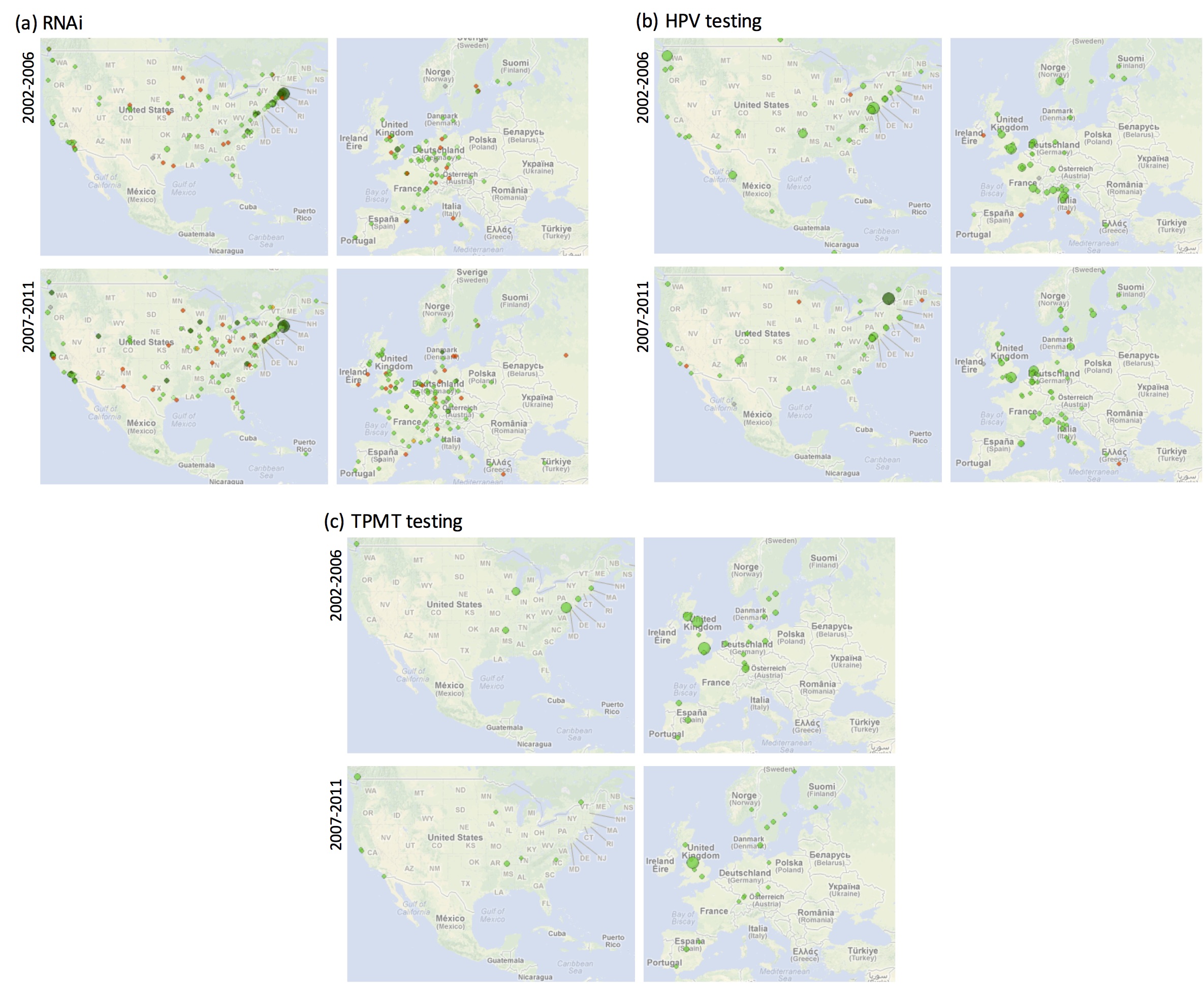}
\centering
\caption{Areas of highly-cited publications for the 2002-2011 period for RNAi (a), HPV (b) and TPMT (c) testing technologies. Nodes are coloured dark green (red) when the difference between the observed number of top-cited publications and the expected one is positive (negative) and statistically significant ($p<0.05$), light green (orange) otherwise. \newline\textit{Source: Authors' elaboration on the basis of WoS data.}}
\label{fig:geopub}
\end{figure}

For the three case-studies of this article, we used overlays projecting publication data corresponding to 5-year time windows.\footnote{The time window does not affect the results in this case. Similar dynamics are observed by using narrower or broader time windows (e.g.\ 3-year, 7-year). It is worth noting that in order to exploit all the available data up to the year 2011, the first time window for RNAi is 1998-2001, which includes four years of observations instead of five years. RNAi can indeed be observed starting from year 1998 when this gene silencing mechanism was discovered \citep{Fire1998}.} Using constant time windows across the case-studies it is possible to capture the differing pace at which the selected technologies have emerged in terms of (WoS) publications. \Figref{geopub} depicts the results of the geographical overlay mapping during the period 2002-2011 (the entire set of maps can be accessed at \url{http://dx.doi.org/10.6084/m9.figshare.1501068}).\footnote{These maps can be also generated by using patent data \citep[see][]{Leydesdorff2012a}.} 

In the case of RNAi (\Figref{geopub}a), the mapping shows the emergence of urban areas producing highly-cited articles on the East Coast of the US near Baltimore and New York, and, especially, in the area of Boston. The RNAi mechanisms were discovered by groups of scientists working in Massachusetts \citep{Haussecker2008}. However, while the maps in the subsequently years report the emergence of many sites all round the world, Massachusetts has persisted as the main area producing highly-cited scientific articles --- this is revealed by the size of the node --- and as one of the hubs for RNAi biotech firms. 

For the case of HPV testing technologies (\Figref{geopub}b), these maps identify, over the entire observation period, European urban areas producing highly-cited scientific knowledge more frequently than expected in London, Paris, and Amsterdam. New areas have also started to appear both in the north (nearby Copenhagen, Helsinki, and Jena) and south of Europe (nearby Barcelona, Bologna, and Turin) since the mid-1990s. The US sites contributing to this emerging technology with highly-cited articles are mainly located on the coasts, specifically in the area of Washington D.C., Baltimore, New York, and Boston, for the East Coast, and near San Francisco and San Diego for the West Coast. Georgetown University and the private company Digene in the area of Washington D.C. have played a key role for the development and the adoption of a test for the HPV detection in cervical cancer screening \citep{Hogarth2012a}. The maps also reveal the rise of new sites producing highly-cited scientific knowledge in South America (e.g.\ near Sao Paulo, Buenos Aires) during the last ten years of observation. 

This geographical mapping for TPMT testing technology (\Figref{geopub}c) locates, at the beginning of the observation period, highly-cited articles in the urban areas of Rochester and Memphis (US) as well as Sheffield (UK). However, as for HPV testing technology, new sites have started to appear since 1997 close to the East and West coasts (e.g.\ areas of Washington D.C., Boston, San Diego, San Francisco), across the UK (e.g.\ near London, Glasgow, Edinburgh) and Europe (e.g.\ Berlin and Madrid).
This analysis shows how overlay mappings can provide the policy-making process with relevant inputs such as the list of geographical areas where main advancements for a given emerging technology were achieved and areas that are 'unexpectedly' listed. It may also inform on areas that persistently contribute over time to the development of the emerging technology as well as on new emerging ones. 

\textit{De facto} coordination arrangements are also revealed. For example, the maps for HPV and TPMT testing technologies identify a number of urban areas in which highly-cited publications concentrate as located in developing countries. A further analysis of the collaboration networks (see below) revealed co-authorship links between these areas and others leading the advancements of these emerging technologies in the developed countries. These collaborations may have provided developing countries with the access to critical capabilities and resources to produce novel and high-quality knowhow, but perhaps indicating some crucial contribution (e.g.\ a genetic resource or patient population) needed by researchers in developed countries too that was helpful. 

It is important to note some limitations of the above-discussed approaches. Firstly, the geographical information reported in publication (and patent) data may not reflect the locations where the research was conducted. Secondly, while the overlays built at city-level provide high granularity, they can represent sites located in the same urban area as two different nodes. For example, in the case of HPV testing technology, Silver Spring (a suburb of the US capital) was considered as a node different from Washington D.C.. These limitations represent directions for future development of these techniques.

\subsection{Tracing emergence in the social space}
The structure of the collaborative relationships among the actors surrounding emerging technologies and their dynamics play a critical role in the emergence process \citep[e.g.][]{Latour1993, Powell1990}. These connections are channels through which actors gain access to and mobilise knowledge, resources, and power. Networks of agents affect and are affected by emerging technologies, i.e.\ actors create social structures over the emergence that both enable and constrain their actions \citep{Giddens1984}. By using co-authorship data \citep[e.g.][]{Crane1972, Glanzel2004,Wagner2008}, the dynamics across this relevant space of emergence can be roughly traced --- however, it should be noted that many collaborations are not formalised into co-authorship relations \citep{Laudel2002}. 

\begin{figure}[h]
\includegraphics[width=16cm]{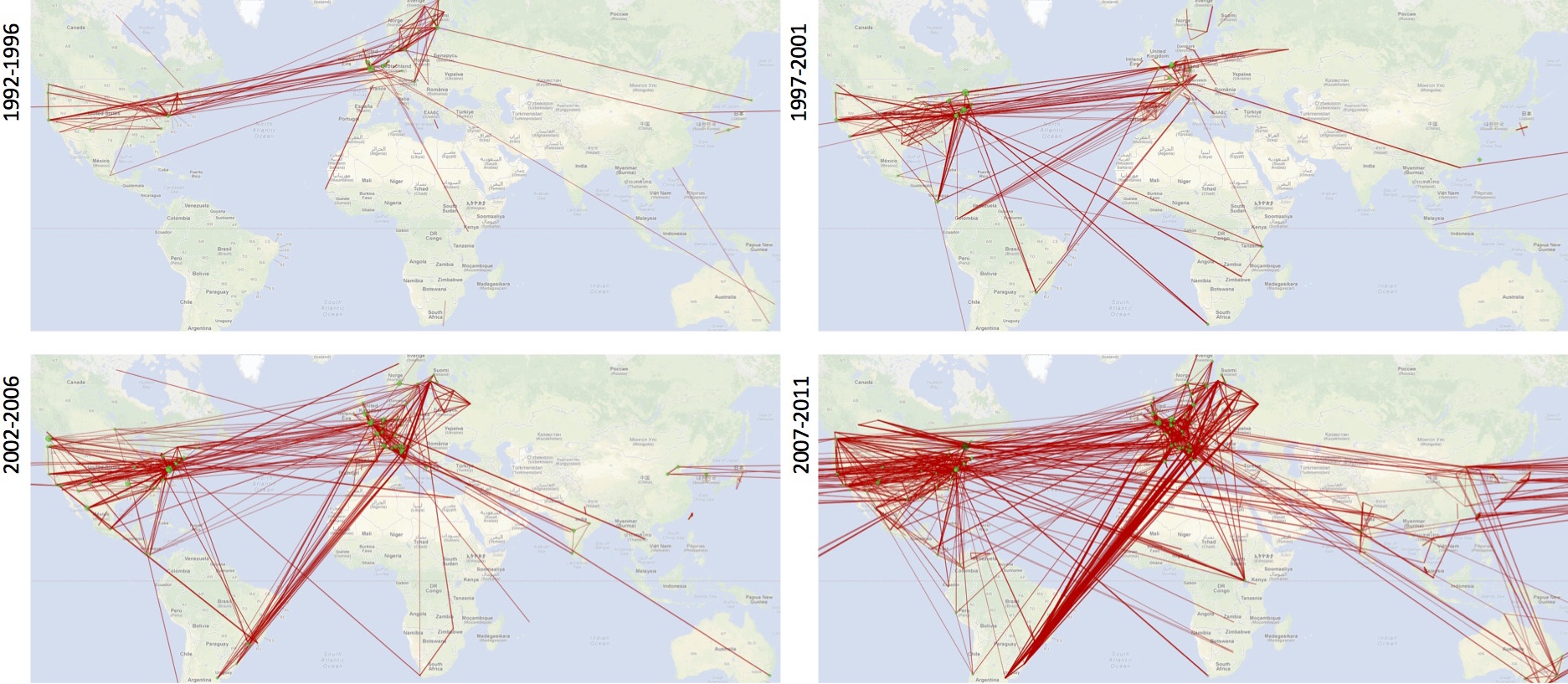}
\centering
\caption{Overlay of the co-authorship network at city-level for HPV testing technologies in the 1992-2011 period. The size of nodes is proportional to the $log_2$ of the number of scientific articles (plus one) that the given city published. \newline\textit{Source: Authors' elaboration on the basis of WoS data.}}
\label{fig:geosoc}
\end{figure}

The application of overlay mapping in the case of the social space is relatively limited because one would need to generate a 'global' collaboration network (at the organisation or author levels) that would function as basemap for the projection of overlays. Generating such a basemap is, however, particularly challenging. Differently from geographical and cognitive basemaps, the 'global' collaborative network is likely to be highly unstable over the observation period (e.g.\ actors joining/leaving the network) notwithstanding the considerable effort required to disambiguate actors' names. While the development of this basemap represents an important direction for future research, here we focus on the type of intelligence about emerging technologies that can be generated by projecting social structures over the geographical space \citep{Leydesdorff2011}. For example, \Figref{geosoc} shows the collaboration networks at city-level for the HPV testing case-study. In this map, nodes are cities and the linkages between nodes are traced by using co-authorship data. Investigating the dynamics across these maps may provide informative perspectives that are derived by combining the geographical and social spaces. Examples of key empirical questions the overlay mapping approach may help to address in the policy-making process are: Where does an emerging technology arise? Does the collaboration network cluster in specific areas? Does the given emerging technology spread across cities, regions, and countries and, if yes, through which (collaboration) channels --- which types of grant were important, which size of institutions, or types of knowledge brokers?

The collaborative network of HPV testing technologies, for example, discloses three relevant dynamics of the emergence process. First, strong collaborative activity between the US (especially the areas of Washington and New York) and Europe (initially Germany) can be observed. This technology indeed started to emerge after a scientist, Harald zur Hausen, at the German Cancer Research Centre first proved that HPV infections are strongly associated with the development of the cervical cancer \citep{ZurHausen1987}. This discovery subsequently found important application in the US where extensive screening programs for cervical cancer were already in place and a potential market for new tests was attractive for entrepreneurs. Here, a small US biotech company, Digene, developed and marketed a series of FDA-approved HPV tests, now recognised as significant milestones in the evolution of screening technologies for identification of patients at risk of cervical cancer. A second dynamic that the mapping reveals is the move, in the last ten years of observation, towards an increasing involvement of developing countries (e.g.\ Brazil, India) in the research networks. Cervical cancer in these countries is a significant social burden and HPV testing was potentially a new relevant approach for cervical cancer screening in these developing countries. A third dynamic is the globalisation of research on HPV testing technologies can be observed across the entire period as revealed by the density of the network of relationships across cities. The latter dynamic is also observed in the case of RNAi that diffuses all round the world, even more rapidly than HPV testing technology (the maps of the three case-studies for the entire observation period can be accessed at \url{http://dx.doi.org/10.6084/m9.figshare.1501086}). 

Building on these cross-cuttings of the geographical and social spaces, one can focus more attention on the social dynamics by looking at the structure of the web of relationships composing the network at a lower level of analysis such as the organisation-level. The network can be explored with algorithms that identify cohesive groups of organisations as well as public and private players occupying key positions. Social network analysis provides a broad range of measures that may support a quantitative assessment of the changes occurring over time in the analysed network.\footnote{It is worth noting that the interpretation of some of these measures is difficult because they are size dependent and networks featuring in emerging technologies show relatively fast growth.} In this regard, the analysis of the structure of co-authorship network at the organizational level for the HPV testing technology case-study (\Tabref{networkdesc}),\footnote{We use The Vantage Point software to harmonise organisations' names.} revealed the emergence of a giant component as well as an increasing average path length of the network. A detailed examination showed Digene occupying a strong and influential position within this network by collaborating with major institutions in the field of cervical cancer screening (e.g.\ National Cancer Institute, a public sector sponsor, and Kaiser Permanente, a large healthcare service provider and early test adopter). This eventually allowed Digene to influence the adoption of the test, for example by working with clinicians involved in the development of medical guidelines \citep{Hogarth2012a}. In other words, while Digene's product development activity was regulated by the FDA, for example, Digene was affecting the developments and dynamics the use of cervical cancer screening by researchers and practitioners.

\setlength{\tabcolsep}{10pt}
\renewcommand{\arraystretch}{1.2}
\begin{table}\footnotesize
	\caption{\label{tab:networkdesc}Descriptive statistics of the co-authorship network at organizational-level for HPV testing technologies.}
	\centering
   \begin{tabular}{lccccccc}

	\hline\hline
	Time window               & 1992-1996   & 1997-2001   & 2002-2006   & 2007-2011   \\
	\hline
	Nodes                        	& 173         & 265         & 471         & 816         \\
	Ties                           	& 223         & 476         & 980         & 2075        \\
	Density                       	& 0.01        & 0.01        & 0.01        & 0.01        \\
	Avarage path length    & 3.4         & 2.8        & 4.4        & 4.4        \\
	Isolated (\%)                & 43 (25\%)   & 47 (18\%)   & 55 (12\%)   & 83 (10\%)   \\
	Components (min. 3 nodes)                   & 11          & 16          & 23          & 40          \\
	First largest comp. (\%)  		& 62 (36\%)   & 83 (31\%)   & 239 (51\%)  & 504 (62\%)  \\
	Second largest comp. (\%) 	& 8 (5\%)     & 12 (5\%)    & 14 (3\%)    & 9 (1\%)     \\
	Degree, Mean (Std.Dev.)       & 2.6 (3.0) & 3.6 (4.8) & 4.2 (4.7) & 5.1 (5.7)  \\
	\hline\hline
	\multicolumn{5}{l}{\footnotesize \textit{Source: Authors' elaboration on the basis of WoS data.}}
    \end{tabular}
 \end{table}

The above discussed examples that combined the geographical and social spaces of emergences and moved across units of analysis, in this case from the city to the organisation level, provides indication of the flexibility and diverse granularity of overlay mapping in generating intelligence on the constellations of actors involved in the emergence process, structure of the relationships among these actors, key actors and collaborations shaping the emergence as well as main channels where knowledge and resources may flow.

\subsection{Tracing emergence in the cognitive space}
As new technologies emerge, epistemic developments occur in terms of discoveries, novel theories, or changes in technical developments such as experimental systems, materials, methods, and instrumentation \citep{Rheinberger1997,Joerges2002}. These dynamics can be traced across the cognitive space by creating overlays of publications on basemaps of science that can be defined at different levels of analysis \citep[e.g.][]{Klavans2009,Waltman2012}.

\begin{sidewaysfigure}
\includegraphics[width=20cm]{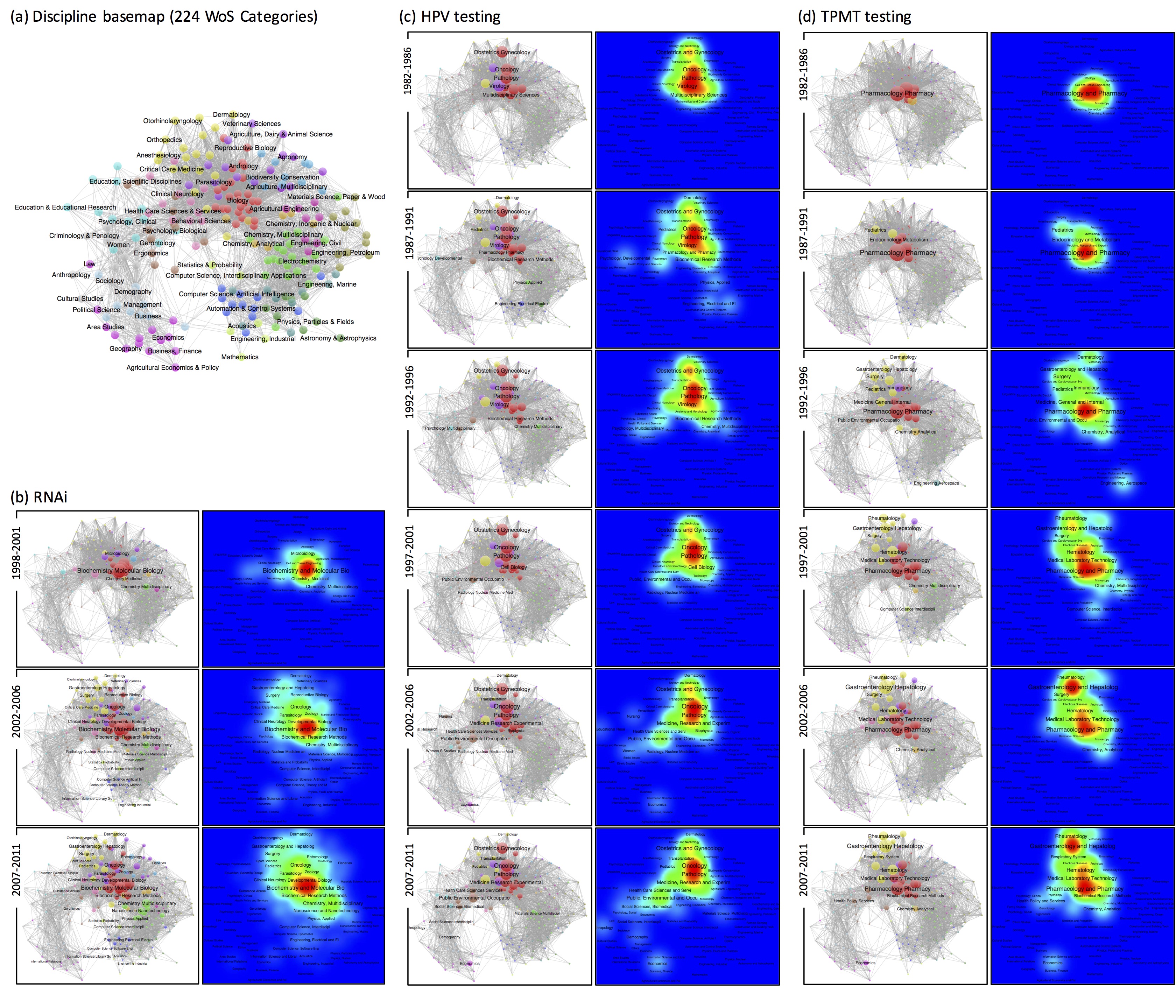}
\centering
\caption{Discipline (WoS Categories) basemap (a) and overlay mapping of RNAi (b), HPV (c) and TPMT (d) testing technologies.  \newline\textit{Source: Authors' elaboration on the basis of WoS data.}}
\label{fig:science}
\end{sidewaysfigure}

The publishing activity related to three case-studies can be, for example, projected across the map of science defined by the 224 WoS Categories \citep[]{Rafols2010}. In this map, each node is a WoS Category that can be assumed as proxy of a scientific subdiscipline. The projection (overlay) makes a node's size proportional to the number of publications related to the given technology that were published in the given discipline the node represents. The different colours of nodes represent different clusters of disciplines. \Figref{science} depicts the projections of publications related to the three case-studies. We report the map representing the structure of science (left) --- the strength of each linkage is proportional to the similarity of citations patterns of two WoS Categories --- and the heatmap version (right), which shows relative density of publications in a given area of the basemap.\footnote{The visualisations of cognitive maps were produced by using VOSviewer 1.5.4 \citep{Eck2010}.} This combination provides an intuitive visualisation of the diffusion process of emerging technologies.

As for the previous analyses, we used overlays projecting the publishing activity according to 5-year time windows. While these maps show the rapid diffusion of RNAi technology as represented by publications in journals read by many disciplines such as molecular biology, oncology, biomedical research, and chemistry, the overlays of HPV and TPMT testing technologies reveal different patterns of diffusion. HPV testing technology diffuses from basic research in oncology, pathology, and virology disciplines towards issues related to the public health. We interpret this dynamic as a representation of the extensive and ongoing debate on the practices adopted for the screening of the population. The debate has been focused on the suitability of HPV testing technology as first an adjunct to, and latterly a substitute for, the widely adopted Pap test \citep{Hogarth2012a}.
 
TPMP testing technology diffuses from the basic research in pharmacology towards clinical disciplines such as gastroenterology and dermatology. Publication activity seems to equally spread in gastroenterology and dermatology disciplines during the 1992-1996 period. Yet, in the subsequent years, the volume of publications shrinks from dermatology area while continuing to grow in gastroenterology. This raises questions about the degree to which the use of TPMT testing technologies use has been contested in the different communities such as gastroenterologists, rheumatologists, and dermatologists.
 
So far we have indicated in several places that the mapping techniques we have presented often raise questions that then require further investigation that analysts may want to pursue for informing the policy-making process. For example, we followed the aforementioned dynamic by conducting an additional analysis of the content of the scientific articles published by these expert communities. One potential conclusion may be that fewer publications in dermatology suggests failure to adopt TPMT testing. However, this is not the case. More fine-grained analysis of journal article titles and abstracts showed that while the community of dermatologists broadly accepted the validity of TPMT test, this test is still highly contested in the communities of rheumatologists and, especially, gastroenterologists. This more intense debate is captured by the cognitive mapping approach. It is also testified by the strongly worded titles of scientific articles and presence of a large number of author response/reply or comment articles to other articles that these two communities published.
 
\begin{figure}[h]
\includegraphics[width=16cm]{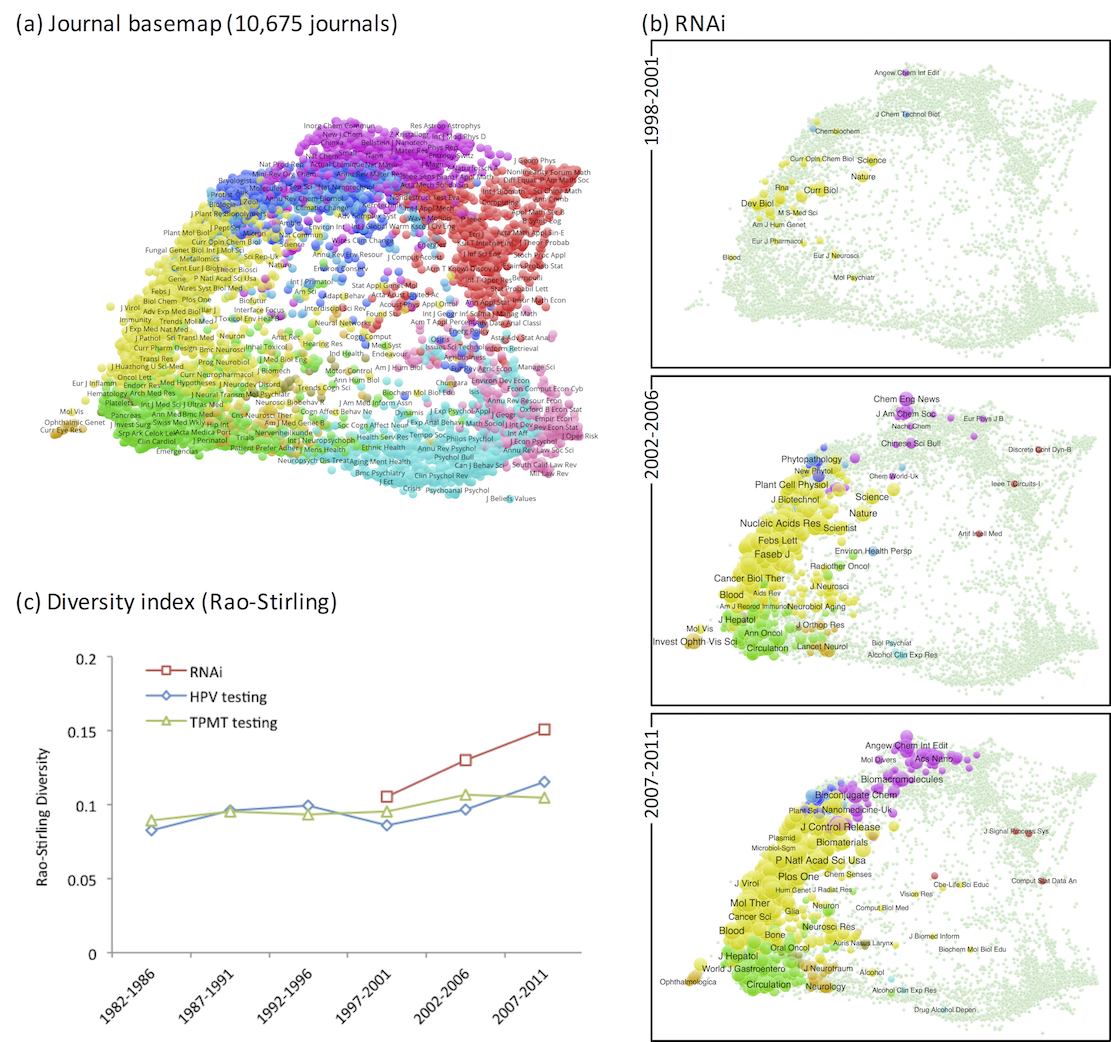}
\centering
\caption{Journal basemap (a), overlay mapping of the RNAi case-study (b), and diversity indexes (c) for the three case-studies.  \newline\textit{Source: Authors' elaboration on the basis of MEDLINE/PubMed.}}
\label{fig:journals}
\end{figure}

A similar cognitive perspective can be built by using a map of which nodes represent academic journals \citep{Leydesdorff2013j}. The map is specifically composed by 10,675 journals (nodes) (\Figref{journals}a) --- the different colours of nodes represent different cluster of journals, i.e.\ groups of journals of which the cross-citation patterns are similar. \Figref{journals}b, for example, illustrates the rapid diffusion of RNAi across this map. RNAi has specifically started to appear in journals in the basic biomedical science and subsequently it has diffused among a variety of discipline-specific journals. The Rao-Stirling diversity index, measured on the set of journals of the map, provides further evidence of this rapid diffusion, especially when the index is compared with the other two emerging technologies on which we focused our analysis (\Figref{journals}c)\footnote{The Rao-Stirling diversity is defined as $\Delta=\sum_{ij}p_ip_jd_{ij}$ \citep{Rao1982, Stirling2007}, where $d_{ij}$ is a disparity measure between two classes $i$ and $j$ --- the categories are in this case journals --- and $p_i$ is the proportion of elements assigned to each class $i$. Disparity is measured in terms of distances on the journal map \citep{Leydesdorff2013j}.} (journal overlay maps of the case-studies for the entire observation period can be accessed at \url{http://dx.doi.org/10.6084/m9.figshare.1501090}).
 
\begin{figure}
\includegraphics[width=16cm]{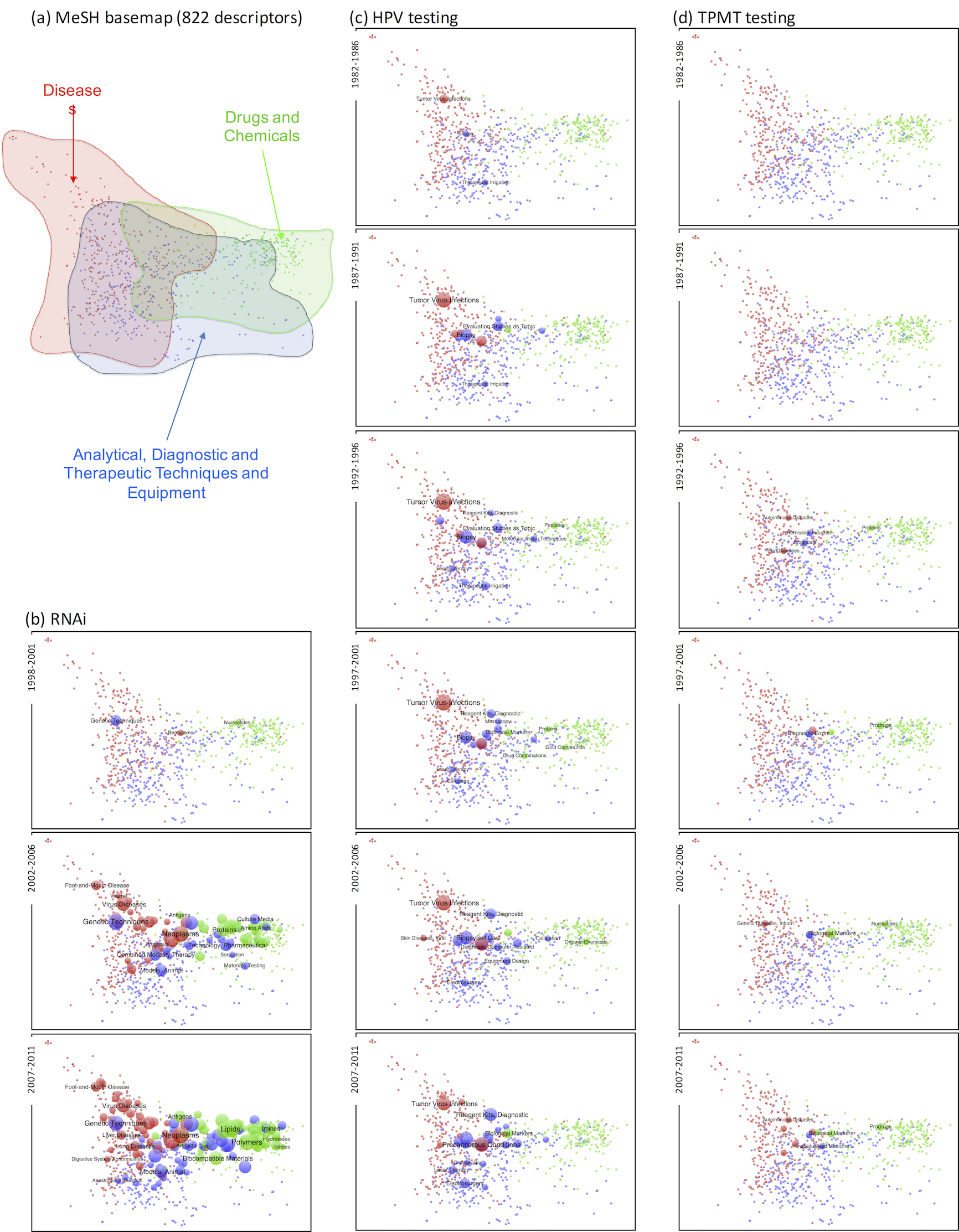}
\centering
\caption{MeSH basemap (a) and overlay mapping of RNAi (b), HPV (c) and TPMT (d) testing technologies.  \newline\textit{Source: Authors' elaboration on the basis of WoS data.}}
\label{fig:mesh}
\end{figure}

Perspectives on the cognitive dynamics can be also built by using MeSH descriptors --- terms used to characterise the content of scientific articles in medical research. These terms are assigned to articles in MEDLINE/PubMed through an intensive indexing process that is performed by examiners at the National Institute of Health (NIH). The terms are organised in a 16-branch tree, which can reach up to 12 levels of depth.  Drawing from this classification, \cite{Leydesdorff2012} developed a MeSH map by using three branches: "Diseases", "Chemicals and Drugs", and "Analytical, Diagnostics and Therapeutic Techniques and Equipment". The first two levels of the tree were selected to build the basemap. The map is specifically composed by 822 MeSH terms (nodes) of which linkages reflect the (cosine) similarity according to co-occurrence of these terms in scientific articles. Each branch is marked on the map with a different colour: "Disease" is coloured red, "Chemicals and Drugs" green, and "Analytical, Diagnostics and Therapeutic Techniques and Equipment" blue (see \Figref{mesh}a). Similarly to previous approaches, the publishing activity characterising a given emerging technology can be projected on this map to trace dynamics across three branches of the MeSH tree. 

This approach, applied to the three case-studies, revealed different dynamics. RNAi, in line with previous results, 'globalises' across the set of the MeSH terms thus affecting many areas of the represented branches (\Figref{mesh}b). By contrast, HPV testing technology diffuses from the "Diseases" branch, specifically from "Tumor Virus Infections", into the "Analytical, Diagnostics and Therapeutic Techniques and Equipment" branch and eventually across the "Chemicals and Drugs" area. Yet, interestingly, in the last time window (2007-2011 period) scientific articles on HPV testing technology concentrate in the techniques and equipment area (\Figref{mesh}c). This may reflect the efforts in developing competing HPV testing technologies. Results also show the specialisation of the TPMT case-study in more limited areas of the map, reflecting its narrow range of application (namely a handful of clinical niches) (\Figref{mesh}d).

Tracing the patenting activity of emerging technologies provides additional perspectives on the cognitive dynamics of the emergence process given the diverse incentives featuring in the creation process of scientific publications and patents. Scholars have developed techniques also to trace the dynamics of the patenting activities \citep[e.g.][]{Kay2014, Schoen2012}. The nodes of these maps are technological classes that, as in the case of previous maps, are linked by cross-citation (cosine) similarity \citep{Leydesdorff2014,Leydesdorff2015}.

\begin{sidewaysfigure}
\includegraphics[width=18cm]{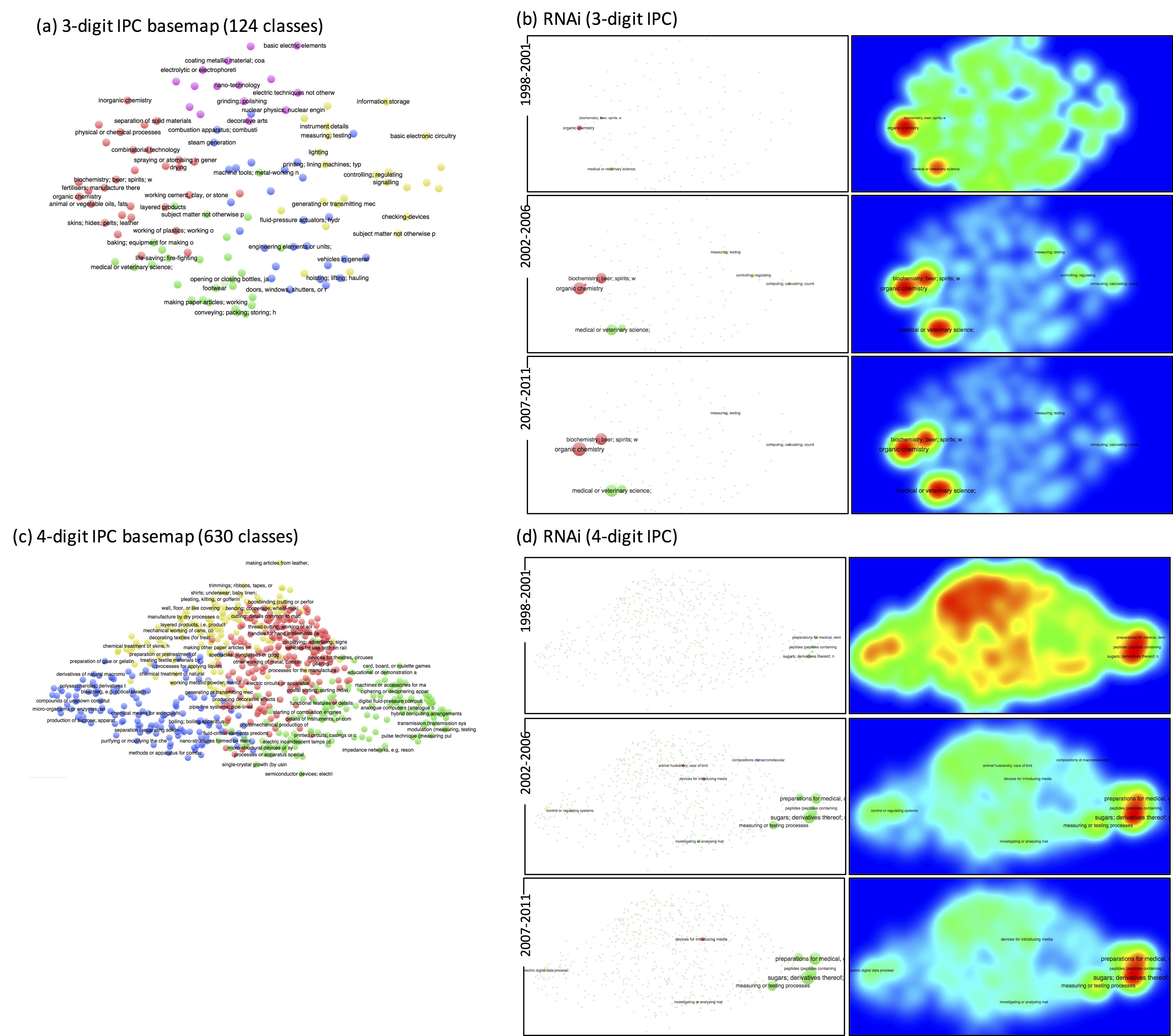}
\centering
\caption{IPC basemap at 3-digit (a) and 4-digit (c) and overlay for RNAi case-study (b,d).  \newline\textit{Source: Authors' elaboration on the basis of USPTO data.}}
\label{fig:patent}
\end{sidewaysfigure}

\Figref{patent}, for example, depicts the overlays of RNAi patenting activity on the patent map based on technological areas as defined by the International Patent Classification (IPC). One can trace the dynamics in this space by moving across different levels of the classification (e.g.\ 3-digit, 4-digit). The patent map visualisation revealed the patenting activity of RNAi focused in specific areas of the technological space such as biochemistry, organic chemistry, and medical science. While one would expect to observe intense patenting in these technological areas given the nature of RNAi, this also reveals how one may need to increase, from case to case, the granularity of the mapping (i.e.\ lower levels of IPC classes) to build informative and interpretative perspectives on the observed emerging technology.

Mapping technological emergence in the cognitive space may reveal a number of important dynamics. These include the directions of diffusion of the given emerging technology across the key knowledge areas involved in emergence, how those areas may integrate or compete, in which domains actors' knowledge production processes are positioned, or, for medical innovation which diseases, a technology is addressing and the type of techniques and molecules these efforts are associated with.

\section{Discussion}
In cases of the emergence of novel technologies, all actors and particularly policy makers have very incomplete knowledge, not knowing, for example, where the boundaries of the technology are, in which direction it is moving, and how it could move forward. In the face of uncertainty, ambiguity, or ignorance \citep{Stirling2007}, the emergence process should be investigated and analysed with strategic intelligence tools in order to support a more informed policy-making process. We have shown how scientometric overlay mapping can support this process by generating intelligence input across the geographical, social, and cognitive spaces (and combinations of these) of emergence. Overlay mapping can be somewhat conceived as a monitoring system for emerging technologies that by tracking emergence and synthesising the complex information in relatively accessible visualisations can potentially favour the development of policy instruments of a higher scope, speed, accuracy, and reliability \citep{Nightingale2003}. 

The use of these techniques as a strategic intelligence tool is, however, dependent on a set of choices that the analyst makes. Although some choices may seem purely technical (e.g.\ level of aggregation), in practice they have an interpretative component that may also have important implications for the type of patterns one can observe (e.g.\ the granularity of results). Firstly, a key preliminary choice is to identify the boundary of emerging technologies, i.e.\ the delineation of the corpus. For analysts, deciding what to include and what to exclude in the analysis is often a problematic exercise. In the case of RNAi, for example, one type of delineation portrayed this technology as already having reached a mature phase, whereas a broader definition suggested RNAi in a phase of rapid growth (\Figref{delineation}). Multiple interactions with experts in the field of the emerging technology under study therefore assume a critical role in this process.

Secondly, the analyst has to identify appropriate data sources. It is clear that using bibliographic data from publications and patents yields different types of information. Yet, those are only one form of research outputs. Many other possible dimensions such as products, services, changes in healthcare outcomes, are not easily accessible for this mapping process. This implies that some emerging technologies will be well represented in certain datasets and not others. This may also be the case for certain groups of actors --- for example, academic organisations may publish more and patent less than private organisations, who may not have as high a propensity to publish papers as they do to apply for patents. Consideration of available datasets places a considerable onus on the analyst to find appropriate sources and not to over-interpret limited data from less appropriate sources. 

Studying medical innovations in an IP dominated industry, where regulatory and peer community pressures encourage publication, is a considerable advantage from the point of view of data access and availability. However, other contexts may not be so well provisioned, with implications for the utility of the approaches discussed here. The extension of overlay mapping approach to other types of data sources represents an important challenge for future research especially when one considers the increasing attention towards the use of 'big data' and altmetrics \citep[e.g.][]{Thelwall2013,Bornmann2015}. Social media data may yield interesting information. Yet these so-called 'altmetrics' techniques may be more sensitive to fads and hype than in the databases where the generation of outputs requires more effort (e.g.\ publications in WoS, MEDLINE/PubMed, and patent applications).

Thirdly, the selection of the elements to be analysed from the records of the databases as well as the categories into which the elements are assigned may have a significant impact on the resulting analyses. From a patent record, for example, one can extract information about inventors, firms, technology classes, or location, which, as discussed, provide insights on the social, cognitive or geographical spaces of emergence, respectively. In the case of categories one needs to decide the level of aggregation (the granularity of the description) as well as the type of classification --- i.e.\ whether a predetermined, top-down typology is used (e.g.\ the MeSH descriptor) or an emerging, bottom-up taxonomy. 

Ideally, the choices of databases, elements, and categories can be informed by conceptual and theoretical frameworks. However, from a policy analyst's standpoint, it may be also critical to monitor technological emergence even when there is lack of an explicit understanding of conceptual frameworks used --- policy is needed under conditions of incomplete knowledge. Yet, even in these cases in which there is no explicit adoption of a conceptual framework, the choice of specific elements and categories in the analysis is privileging certain understanding over others, and the implicit assumptions that drive the analysis need to be made clear to ensure the maps are interpreted within an appropriate context. It is all too easy to misread a map. For example, looking at collaborative networks of individual scientists implicitly places more attention on the social capital as a key factor in the emergence process. If one looks at the disciplinary position of the technology it is likely to be assuming that integration of disparate knowledge is relevant as, for example, it has been perceived to be the case of nanotechnologies \citep[e.g.][]{ Porter2009} and in RNAi \citep{Leydesdorff2011}, but not necessarily in HPV or TPMT testing technologies. In the absence of a clear understanding of the technology, one is advised to explore the phenomenon by using several perspectives, because one does not know in advance which one may turn out to be useful for understanding the relations of the emergence in place, or because the areas in which the action is occurring are shifting over time. 

Emerging technologies do not conform to established bodies of knowledge. They cut across pre-existing organisational and institutional units, challenging established managerial and policy practices. As a result, one key demand from analysts and decision-makers is a description of the types of interdisciplinarity or convergence evident in emergence \citep{Schmidt2007}, often related with specific visions and expectations \citep{Roco2002,Beckert2007}. In this regard, building multiple perspectives on the process of emergence requires looking across databases where emerging technologies 'tumble' with different representations that stress different attributes --- such as geographical addresses, WoS Categories, patent classes.

The databases are retained and organized in different contexts with relative institutional rigidities. For example, relating patents to publications in terms of 'non-patent literature references' requires professional skills and cannot be done on a large scale without substantive investments. Thus, while the mapping and overlay techniques allow one to use the same or similar search strings across these databases more research is required to increase the integration of datasets. This can already be seen in attempts by US funders, such as the National Institutes of Health (NIH), to link research grants, patents and publications.\footnote{See, for instance, \url{http://projectreporter.nih.gov/reporter.cfm}}  

In summary, while overlay maps are a potentially useful tool for informing decision-making, the adequacy of this tool for mapping emerging technologies is quite sensitive to a variety of choices (such as boundary of technology, data sources used, elements and level of granularity of analyses). Because the maps are generated during emergence, there is a high degree of uncertainty and ambiguity regarding the choices to be made. As a result, these maps are not expected to be accurate or even very reliable -- they are just heuristic pictures of shapes yet in the making. The purpose of using multiple dimensions to investigate emergence is to help in spotting robust insights against patterns that are the result of analytical artifacts.

\section{Conclusions}

In this article, we have examined the use of scientometric overlay mapping as a tool of strategic intelligence for the governance of emerging technologies. To do so, we have elaborated a synthesis of different mapping approaches, and applied these on three case-studies of emerging technologies in the medical domain to illustrate the variety of 'intelligence' inputs for the policy-making process that they can generate. Relevant dynamics of emerging technologies can be illustrated across the geographical, social, and cognitive spaces, i.e.\ in terms of geographical distribution of knowledge processes, collaborative interactions among geographical areas and organisational actors, and scientific and technological domains emerging or involved. Overlay mapping can also crosscut sources of data and move across units of analysis (organisations, cities, disciplines, technologies and sub-fields). This flexibility and diverse granularity favour the comparison of the results from different policy-making contexts.

The three case-studies of emerging technologies we examined (RNAi, HPV, and TPMT testing) enabled us to elaborate an integrative synthesis of different overlay mapping techniques and the types of perspectives on emergence these can generate. In the case of HPV and TPMT testing technologies, we showed how different facets and dynamics of the emergence process can be revealed by moving across the geographical and social spaces from the city-level to organisation-level units of analysis. The maps also enable the user to compare different technologies and alternatives in terms of strengths and weaknesses at the portfolio level. As in the case of RNAi, the mapping can reveal cases of emerging technologies initially concentrated at a few places with major players hosting the sets of relevant capabilities, competences, and network relationships, and then diffusing across a number of actors according to preferential attachment mechanisms \citep{Leydesdorff2011}. On this 'journey' of the emerging technology, changes in the involved domains of science and technology --- especially in the case of RNAi, the diffusion of laboratory research tools in other specialties and disciplines --- can also be revealed with the cognitive mapping.

Using overlay maps has some potential problems that need to be addressed during the study, namely: the delineation issues associated with emerging technologies, the need of some qualitative background on the studied technology for the interpretation and refinement of the mapping, the limitations of the data that can be used (mainly publications and patents). If these problems prove manageable, overlay mapping can function as a heuristic tool that can reveal key trends and allow stakeholders to bring relevant data to theoretical and policy debates. Maps throw new light on relationships and raise additional questions for analysts and decision-makers to address, thus opening up the space of discussion. The resulting perspectives on emergence, according to the set of choices an analyst takes, may help debate in a timely manner the directions of further investigation as well as feeding into political discourse about stumbling blocks ahead and possible openings in the landscapes.

\section*{Acknowledgements}
The authors acknowledge the support of UK Economic and Social Research Council (award RES-360-25-0076 -  \href{http://www.interdisciplinaryscience.net/mdetp}{"{\color{blue}Mapping the Dynamics of Emergent Technologies}"}) in the development of the case-studies. The authors are also grateful to the US National Science Foundation (award \#1064146 - \href{http://www.interdisciplinaryscience.net/scisip}{"{\color{blue}Revealing Innovation Pathways: Hybrid Science Maps for Technology Assessment and Foresight}"}) for the support in the development of the mapping techniques. The findings and observations contained in this paper are those of the authors and do not necessarily reflect the funders' views. The authors are grateful to Paul Nightingale, Stefan Kuhlmann, and Peter Stegmaier, and the two anonymous reviewers of the Journal of the Association for Information Science and Technology for their suggestions and comments on earlier versions of this paper. Previous versions of this paper were presented at the 2013 ISSI Conference, 2013 EU-SPRI Conference, 2013 Atlanta Conference, 2013 DRUID Conference, and "2013 Data Science, Tech Mining and Innovation" workshop in Manchester.

\singlespace
\bibliographystyle{apalike}

\end{document}